\documentstyle[floats,aps,aipbook]{revtex}
\input psfig.sty
\begin{document}
\def \ite{{\it et al.}}
\title{$B$ Physics: Theoretical Aspects}
\author{Jonathan L. Rosner \\
\vspace{0.4in}
EFI 95-33; hep-ph/9506316
\vspace{-0.7in}}
\address{Enrico Fermi Institute and Department of Physics \\
University of Chicago, Chicago, IL 60637}
\maketitle
\begin{abstract}
Some aspects of $B$ physics relevant for experiments in hadron colliders are
discussed.  These include the determination of parameters of the CKM Matrix and
confirmation of its role in CP violation, studies of mixing of nonstrange and
strange $B$ mesons, lifetimes of hadrons containing $b$ quarks, the use of
``same-side'' tagging of neutral $B$ mesons via correlations with charged pions
through fragmentation or resonances, and the determination of CKM phases
through the study of decays of $B$ mesons to pairs of light hadrons.
\end{abstract}

\section{INTRODUCTION}

While the first states containing the $b$ quark, the $\Upsilon$ resonances,
were discovered in hadronic interactions at Fermilab \cite{ups}, the study of
$B$ mesons for many years has been largely the province of $e^+ e^-$
colliders. With the reconstruction of large numbers of exclusive $B$ decays by
the CDF Collaboration \cite{Bed}, that situation now has the potential to
change.  The present talk outlines some of the ways in which hadron colliders
can exploit their innate advantage of large $B$ production rates to provide
information complementary and in many cases superior to that available in
electron-positron collisions.

We begin in Sec.~II with a discussion of the Cabibbo-Kobayashi-Maskawa (CKM)
matrix \cite{cab,KM} and its role in CP violation as currently observed in the
neutral kaon system.  The role of $B$ mesons in checking this picture is
outlined in Sec.~III.  We comment specifically on $B - \bar B$ mixing in
Sec.~IV and on systematics of lifetime differences in Sec.~V.  The
identification of the flavor of a neutral $B$ meson by means of correlations
with charged pions produced nearby in phase space, via fragmentation and/or
resonances, is reviewed in Sec.~VI.  Decays of $B$ mesons to pairs of hadrons
containing light ($u,~d,~s$) quarks can provide information on phases of CKM
elements, as noted in Sec.~VII.  We mention some specific issues associated
with studies in hadron colliders in Sec.~VIII, and conclude in Sec.~IX.

\section{CKM MATRIX AND CP VIOLATION}

\subsection{The Neutral Kaon System}

The mass eigenstates of neutral kaons, $K_S$ and $K_L$, both of which decay to
$\pi \pi$, may be expressed in terms of a complex parameter $\epsilon$:
\begin{equation}
K_{S,L} = \frac{1}{\sqrt{2(1 + |\epsilon|^2)}}
\left[ (1 + \epsilon) K^0 \pm (1 - \epsilon) \bar K^0 \right]~~~.
\end{equation}
The leading candidate for the source of $\epsilon$ at present is a CP-violating
$K^0 - \bar K^0$ mixing term, dominated by the top quark in the loop diagram
for $s \bar d \to d \bar s$ and complex as a result of phases in the CKM
matrix elements.

An analogue of the $K_S - K_L$ system in the absence of CP (or time-reversal!)
violation is provided by two resonant circuits with equal natural frequencies
and damping terms, coupled through a resistor, so that the long-lived
oscillations will be those in which the two systems are in phase with one
another and no energy is dissipated in the resistor.  Emulation of CP violation
is harder.  One may do so using asymmetric exchange of radiated energy between
obliquely polarized antennas through a region in which Faraday rotation is
occurring.  Thus, for example, VLF (very-low-frequency) radio waves are subject
to different attenuation when propagating in the ionosphere from east to west
and from west to east \cite{Davies}.  A more ``table-top'' version of such a
phenomenon could utilize the Hall effect.

\subsection{The CKM Matrix}

In a parametrization \cite{wp} in which the rows of the CKM matrix are labelled
by $u,~c,~t$ and the columns by $d,~s,~b$, we have
\begin{equation}
V = \left ( \begin{array}{c c c}
V_{ud} & V_{us} & V_{ub} \\
V_{cd} & V_{cs} & V_{cb} \\
V_{td} & V_{ts} & V_{tb}
\end{array} \right )
\approx \left [ \matrix{1 - \frac{\lambda^2}{2} & \lambda
& A \lambda^3 ( \rho - i \eta ) \cr
- \lambda & 1 - \frac{\lambda^2}{2} & A \lambda^2 \cr
A \lambda^3 ( 1 - \rho - i \eta ) & - A \lambda^2 & 1 \cr } \right ]~~.
\end{equation}
The third quark family ensures that there will be a non-trivial phase, taken
here to be that of $V_{ub}$.  The elements on and below the diagonal are
determined by unitarity. The phase of $V_{td}$ is appreciable, while those of
$V_{cd}$ and $V_{ts}$ are negligible for present purposes.  These phases allow
the standard $V - A$ interaction to generate CP violation as a higher-order
weak effect.

The four parameters are measured in the following manner \cite{DPF}:

(1) The parameter $\lambda$ is measured by a comparison of strange particle
decays with muon decay and nuclear beta decay, leading to $\lambda \approx \sin
\theta \approx 0.22$, where $\theta$ is the Cabibbo angle.

(2) The dominant decays of $b$-flavored hadrons occur via the element $V_{cb}
= A \lambda^2$.  The lifetimes of these hadrons and their semileptonic
branching ratios then lead to an estimate $A = 0.79 \pm 0.06$, or $V_{cb} =
0.038 \pm 0.003$.

(3) The decays of $b$-flavored hadrons to charmless final states allow one to
measure the magnitude of the element $V_{ub}$ and thus to conclude that
$\sqrt{\rho^2 + \eta^2} = 0.36 \pm 0.09$.

(4) The phase of $V_{ub}$, Arg $(V_{ub}^*) = \arctan(\eta/\rho)$, must be
determined with the help of indirect information associated with contributions
of higher-order diagrams involving the top quark.

The unitarity of V and the fact that $V_{ud}$ and $V_{tb}$ are very close to 1
allow us to write $V_{ub}^* + V_{td} \simeq A \lambda^3$, or, dividing by a
common factor of $A \lambda^3$, $\rho + i \eta ~~ + ~~ (1 - \rho - i \eta) =
1$. The point $(\rho,\eta)$ thus describes in the complex plane one vertex of a
triangle [with angle $\alpha$ in a conventional notation \cite{NQ}] whose other
two vertices are $(0,0)$ (with angle $\gamma$) and $(0,1)$ (with angle
$\beta$).

\subsection{Constraints on $\rho$ and $\eta$}

We obtain a set of constraints on the parameters $\rho$ and $\eta$ in the CKM
matrix in the following manner.  More details may be found in Ref.~\cite{DPF};
we have updated some values.

(1) Charmless $B$ decays lead to the estimate $|V_{ub}/V_{cb}| = 0.08 \pm 0.02$
and hence to the constraint $\sqrt{\rho^2 + \eta^2} = 0.36 \pm 0.09$ mentioned
above.  The limits correspond to the dotted semicircles with center (0,0) in
Fig.~1.

\begin{figure}
\centerline{\psfig{figure=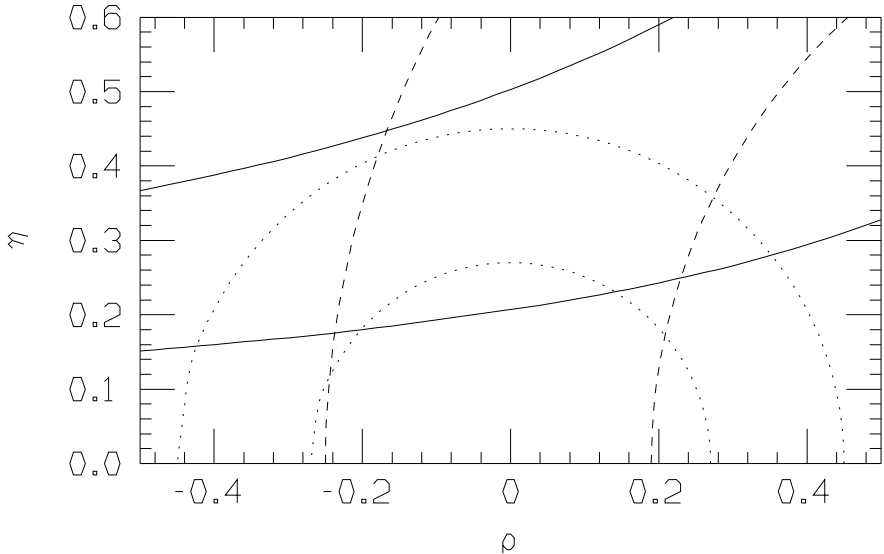,height=3in}}
\caption{Region in the $(\rho,\eta)$ plane allowed by constraints on
$|V_{ub}/V_{cb}|$ (dotted semicircles), $B^0 - \bar B^0$ mixing (dashed
semicircles), and CP-violating $K - \bar K$ mixing (solid hyperbolae).}
\end{figure}

(2) Recent averages \cite{BH} for the $B^0 - \bar B^0$ mixing parameter $\Delta
m_d = 0.462 \pm 0.026$ ps$^{-1}$ and the $B^0$ lifetime $\tau(B^0) = 1.621 \pm
0.067$ ps can be combined to yield $\Delta m/ \Gamma = 0.75 \pm 0.05$.  If
interpreted in terms of a box diagram for $b \bar d \to d \bar b$ (dominated by
the top quark), this value leads to an estimate for $|V_{td}|$ reducing to
$|1 - \rho - i \eta| = 1.03 \pm 0.22$. The corresponding limits are shown as
the dashed lines in Fig.~1.

(3) The parameter $|\epsilon_K| = (2.26 \pm 0.02) \times 10^{-3}$ can be
reproduced with a CP-violating $K^0 - \bar K^0$ mixing term due to the box
diagram for $s \bar d \to d \bar s$.  The contribution of top quarks alone
would be proportional to Im $(V_{td}^2) \sim \rho(1 - \eta)$; with a small
additional charm contribution, one finds $\eta(1 + 0.35 - \rho) = 0.48 \pm
0.20$, whose limits are denoted by the solid hyperbolae in Fig.~1.

Here we have also used the recent average of CDF \cite{CDFt} and D0 \cite{D0t}
top quark masses, $m_t = 180 \pm 12$ GeV/$c^2$, vacuum saturation factors $B_K
= 0.8 \pm 0.2$ and $B_B = 1$, a $B$ meson decay constant $f_B = 180 \pm 30$
MeV, and a QCD correction for the box diagram $\eta_B = 0.6 \pm 0.1$. A large
region centered about $\rho \simeq 0$, $\eta \simeq 0.35$ is permitted.
Nonetheless, it could be that the CP violation seen in kaons is due to an
entirely different source, perhaps a superweak mixing of $K^0$ and $\bar K^0$
\cite{sw}. In that case one could probably still accommodate $\eta = 0$ (see
Fig.~1), and hence a real CKM matrix.  In order to confirm the predicted
nonzero value of $\eta$, we turn to $B$ mesons for help.

\section{ROLE OF $B$ MESONS}

\subsection{Direct $b$ Transitions}

The error in the $b \to c$ CKM element is dominated by theoretical uncertainty
in extracting information from the measured $B$ lifetime and semileptonic
branching ratio.  Even more theoretical uncertainty is associated in extraction
of $|V_{ub}|$ from a small portion of the semileptonic decay spectrum involving
leptons beyond the endpoint for $b \to c \ell \nu$.  In the long run a
measurement of the branching ratio for $B \to \tau \nu$ or $B \to \mu \nu$,
though challenging, will be helpful in extracting the product $f_B |V_{ub}|$.
The expected branching ratios are about $(1/2) \times
10^{-4}$ for $\tau \nu$ and $2 \times 10^{-7}$ for $\mu \nu$.

One should not forget the possibility of eventually getting information on
$|V_{tb}|$ through the decay $t \to W b$, which is expected to have a partial
width of $1.8 \pm 0.4$ GeV for $m_t = 180 \pm 12$ GeV/$c^2$ and $|V_{tb}|
\approx 1$.

\subsection{$B - \bar B$ Mixing}

The error on $|V_{td}|$ extracted from $\Delta m/ \Gamma$ for nonstrange
neutral $B$ mesons is dominated by uncertainty in $f_B$.  In Sec.~IV we shall
mention some ways in which this uncertainty may be reduced, both by direct
calculation and by comparison of mixing in nonstrange and strange $B$ mesons.

\subsection{CP Violation}

Two main avenues for detecting CP violating in systems involving $b$ quarks
involve 1) decays to CP eigenstates \cite{BCP}, and 2) decays to CP
non-eigenstates. In both cases, partial rates for particle and antiparticle
decays are compared, but experimental aspects and interpretations differ.

{\it 1. In decays to CP eigenstates}, one compares the partial rate for a decay
of an initial $B^0$ with that for an initial $\bar B^0$.  The interference of
direct decays (such as $B^0 \to J/\psi K_S$) and those involving mixing (such
as $B^0 \to \bar B^0 \to J/\psi K_S$) gives rise to rate asymmetries which can
be easily interpreted in terms of the angles $\alpha,~ \beta,~\gamma$.  Thus,
if we define
\begin{equation}
A(f) \equiv \frac{\Gamma(B_{t=0} \to f) - \Gamma(\bar B_{t=0} \to f)}
{\Gamma(B_{t=0} \to f) + \Gamma(\bar B_{t=0} \to f)}~~~,
\end{equation}
we have, in the limit of a single direct contribution to decay amplitudes,
\begin{equation}
A(J/\psi K_S, \pi^+ \pi^-) = - \frac{x_d}{1+x_d^2} \sin(2\beta,2\alpha)~~~,
\end{equation}
where $x \equiv \Delta m/\Gamma$.  This limit is expected to be very
good for $J/\psi K_S$, but some correction for penguin contributions is
probably needed for $\pi^+ \pi^-$.  The value $x_d = 0.75 \pm 0.05$ is
nearly optimum to maximize the coefficient of $\sin(2\beta,2\alpha)$.

For the range of parameters noted in Fig.~1, we expect $-0.4 \le A(J/\psi
K_S) \le -0.1$, i.e., an asymmetry of a definite sign, and $-0.47 \le
A(\pi^+ \pi^-) \le 0.3$, i.e., nearly any asymmetry within the possible
limits imposed by the factor $x_d/(1+x_d^2)$.  Thus, the $J/\psi K_S$
asymmetry is likely to provide a consistency check, while the $\pi^+ \pi^-$
asymmetry should be more useful in specifying the parameters $\rho,\eta$
(unless it lies outside the expected limits).

In order to employ this method it is necessary to know the flavor of the
produced $B$ meson.  We shall remark on one possible ``tagging'' method in
Sec.~VI.

{\it 2. In decays to CP non-eigenstates}, the asymmetry is proportional to
$\sin \Delta \phi_{\rm wk} \sin \Delta \phi_{\rm str}$, where $\Delta
\phi_{\rm wk}$ is a difference between weak phases of two interfering
amplitudes, while $\Delta \phi_{\rm str}$ is the corresponding strong
phase shift difference in the two channels.  Such channels might, for
instance, be the $I = 1/2$ and $I = 3/2$ amplitudes for $B \to \pi K$.
Here one is able to compare decays of charged $B$ mesons with those of
their antiparticles, so the identification of the flavor of the decaying
meson does not pose a problem.  On the other hand, $\Delta \phi_{\rm str}$
is generally expected to be small and quite uncertain for the energies
characteristic of $B$ decays.  We shall outline in Sec.~VII some recent
progress in using decays of charged $B$ mesons to provide information on
CKM phases {\em without} necessarily having to observe a CP-violating
decay rate asymmetry.

\section{$B - \bar B$ MIXING}

\subsection{Nonstrange $B$ mesons}

As mentioned, the dominant error in extracting $|V_{td}|$ from $B^0 - \bar B^0$
mixing is associated with uncertainty in $f_B$.  Lattice gauge theory
calculations \cite{LATFB} indicate $f_B \approx 180 \pm 50$ MeV.  A quark model
calculation \cite{QMFB} uses isospin splittings in $D$ and $B$ mesons to
estimate the heavy quark -- light antiquark wave function at zero separation,
obtaining $f_B = 180$ MeV with a negligible experimental error but a
systematic error which is hard to estimate.

Although no direct information on $f_B$ is available, we can
check some of the predictions of the above models for $D$ mesons.
Several experimental determinations \cite{FDS} of the $D_s$ decay constant
have appeared in the last two years, leading to $f_{D_s} \approx 300$ MeV.
The estimate of Ref.~\cite{QMFB} finds
\begin{equation} \label{eqn:SU}
f_D/f_{D_s} \simeq (m_d/m_s)^{1/2} \simeq 0.8 \simeq f_B/f_{B_s}~~~,
\end{equation}
or $f_D \approx 240$ MeV, not far below the present
upper limit of 290 MeV \cite{MkIII}.  Lattice estimates
for the ratio in Eq.~(\ref{eqn:SU}) range between 0.8 and 0.9.

\subsection{Strange $B$ mesons}

The mixing of neutral strange $B$ mesons is expected to involve a ratio $x_s
\equiv \Delta m_s/\Gamma_s \simeq 20$.  Observation of this quantity poses a
challenge but would provide very useful information on $f_{B_s}$, since the
corresponding CKM elements are assumed well known from unitarity: $|V_{tb}|
\approx 1$, $|V_{ts}| = 0.038 \pm 0.003$.  In turn, one would estimate $f_B$
from $f_{B_s}$ using Eq.~(\ref{eqn:SU}).

We show in Table 1 the dependence of $x_s$ on $f_{B_s}$ and $m_t$.  We have
updated a similar table in Ref.~\cite{DPF} on the basis of more recent
information on the $B_s$ lifetime and the top quark mass.  The
dominant source of error on $x_s$ is primarily $f_{B_s}$.
Our value of $f_B$ and Eq.~(\ref{eqn:SU}) imply $f_{B_s} \approx 225$ MeV.

\begin{table}
\begin{center}
\caption{Dependence of mixing parameter $x_s$ on top quark mass and
$B_s$ decay constant.}
\medskip
\begin{tabular}{c c c c} \hline
\null \qquad $m_t$ (GeV/$c^2$)&  168  &  180  &  192  \\ \hline
$f_{B_s}$ (MeV)               &       &       &       \\
150                           &   9   &   10  &   11  \\
200                           &  15   &   17  &   19  \\
250                           &  24   &   27  &   30  \\ \hline
\end{tabular}
\end{center}
\end{table}

The estimate $f_B/f_{B_s} = 0.8 - 0.9$ and an experimental value for
$x_s$ would allow us to tell whether the unitarity triangle
had non-zero area by specifying $|1 - \rho - i \eta|$ \cite{DPF}.
Present bounds are not yet strong enough for this purpose.
(See, e.g., Ref.~\cite{ALBs}, for which the largest claimed
lower bound is $x_s > 9$.)
Assuming that $|V_{ub}/V_{cb}| > 0.06$, one must show $0.73
< |1 - \rho - i \eta| < 1.27$. Taking the $B_s$ and
$B_d$ lifetimes to be equal, and assuming $0.7 < x_d < 0.8$, this will be so if
$16 < x_s < 34$.  An ``ideal'' measurement would thus be $x_s = 25 \pm 3$.

\section{LIFETIMES}

\subsection{Charged vs. Neutral Nonstrange Mesons}

The nonleptonic decay of a $B^0$ to a charmed final state involves the quark
subprocess $b (\bar d) \to d \bar u c (\bar d)$, where the parentheses denote
the spectator quark.  This is to be compared with the decay of a $B^-$,
involving $b (\bar u) \to d \bar u c (\bar u)$.  Here we have assumed the weak
current produces a $d \bar u$ pair.

The corresponding decays for charmed mesons are, for a $D^0$, $c (\bar u) \to
u \bar d s (\bar u)$, and, for a $D^+$, $c (\bar d) \to u \bar d s (\bar d)$.
Now, $\Gamma_{\rm tot}(D^0) \approx 2.5 \Gamma_{\rm tot}(D^+)$, as a
result of different nonleptonic decay rates.  This difference has been
ascribed to many causes \cite{diffs,Blifes}, but almost certainly is due in
part to final-state interactions favoring the non-exotic final state in $D^0$
decays \cite{FSI}.  An exotic state is one which cannot be formed from a
quark-antiquark pair.  In the case of $D^0$ decays, the $u$ and spectator $\bar
u$ can annihilate one another, leaving a non-exotic $s \bar d$ final state.
The $u \bar d s \bar d$ final state in $D^+$ decay is exotic.  The study of
amplitude triangles for such decays as $D \to K \pi$ and $D \to K^* \pi$
indicates that final-state phase differences between the non-exotic $I = 1/2$
and exotic $I = 3/2$ final states indeed can be significant \cite{StoneD}.

It would be surprising if some remnant of this effect were not found in $B$
decays.  The $d \bar u c \bar d$ final state in $B^0$ decay can transform into
a non-exotic $\bar u c$ state through $d \bar d$ annihilation, while the $d
\bar u c \bar u$ final state in $B^+$ decay is exotic. So far, only upper
limits on phase shift differences in channels like $\bar B \to (D \pi,~D
\rho,~D^* \pi)$ have been obtained \cite{HY}.  Such phase differences should be
observable if there really are lifetime differences between charged and neutral
$B$'s and if they can be viewed in terms of final-state interactions.

Bigi \ite~\cite{Blifes} estimate
\begin{equation} \label{eqn:liferatio}
\frac{\Gamma_{\rm tot}(B^0)}{\Gamma_{\rm tot}(B^+)} = 1 + 0.05 \frac{f_B^2}
{(200~{\rm MeV})^2}~~~,
\end{equation}
but their predictions also include a semileptonic $B$ decay branching ratio of
at least 12\%, whereas the experimental value seems to be below 11\% \cite{BH}.
Thus there may be room for some additional enhancement of the hadronic decay
rate. [The subprocess $b \to c \bar c s$, followed by $c
\bar c$ annihilation, can lead to non-exotic states for both charged and
neutral $B$ decays, accounting in part for the discrepancy
between theory and experiment in the semileptonic branching ratio by
enchancing {\em both} charged and neutral $B$ nonleptonic decays slightly.]

The experimental $B^0$ and $B^+$ lifetimes appear to be equal to within
7\% \cite{Bed}.  Greater accuracy is needed in order to test
Eq.~(\ref{eqn:liferatio}) incisively.

\subsection{Baryons}

The charmed baryon $\Lambda_c$ has a remarkably short lifetime:
$\tau(\Lambda_c) \approx \frac{1}{2} \tau(D^0)$.  It was predicted \cite{BLP}
that the $W$-exchange subprocess $cd \to su$, leading to $\Lambda_c(ucd) \to
\Sigma^* (usu) \to$ hadrons, would significantly enhance the $\Lambda_c$ decay
rate, as seems to be the case.  A similar mechanism involving the subprocess
$bu \to cd$ could enhance the $\Lambda_b$ decay rate. Indeed, $\tau(\Lambda_b)
= 1.17 \pm 0.09$ ps \cite{BH}, in contrast to the average meson lifetime $\bar
\tau(B)$, which exceeds 1.5 ps \cite{Bed,BH}.  Bigi \ite~\cite{Blifes} predict
$\tau(\Lambda_c)/\bar \tau(B) \simeq 0.9$, somewhat less of a difference than
observed, but in the right direction.

\subsection{Strange $B$ mesons}

The decay of a $\bar B_s = b \bar s$ meson via the quark subprocess $b (\bar s)
\to c \bar c s (\bar s)$ gives rise to neutral final states which turn out to
be predominantly CP-even \cite{CPeven}.  The mixing of $\bar B_s$ and $B_s$
leads to eigenstates $B_s^{\pm}$ of even and odd CP; the predominance of
CP-even final states formed of $c \bar c s \bar s$ means that the CP-even
eigenstate will have a shorter lifetime.  With $\Delta \Gamma(B_s) \equiv
\Gamma(B_s^+) - \Gamma(B_s^-)$ and $\bar \Gamma(B_s) \equiv [\Gamma(B_s^+) +
\Gamma(B_s^-)]/2$ , Bigi \ite~\cite{Blifes} estimate
\begin{equation} \label{eqn:widthdiff}
\frac{\Delta \Gamma}{\bar \Gamma(B_s)} \simeq 0.18 \frac{f_{B_s}^2}
{(200~{\rm MeV})^2}~~~,
\end{equation}
possibly the largest lifetime difference in hadrons containing $b$ quarks.

One could measure $\bar \Gamma(B_s)$ using semileptonic
decays, while the decays to CP eigenstates could be measured by studying
the correlations between the polarization states of the vector mesons
in $B_s^{\pm} \to J/\psi K_S$.  [For a similar method applied to decays
of other pseudoscalar mesons see, e.g., Ref.~\cite{Nelson}.]

The ratio of the mass splitting to the width difference between CP eigenstates
of strange $B$'s is predicted to be large and independent of CKM matrix
elements \cite{IsiBs,BP} (to lowest order, neglecting QCD corrections which may
be appreciable):
\begin{equation}
\left| \frac{\Delta m}{\Delta \Gamma} \right|
\simeq \frac{2}{3 \pi} \frac{m_t^2 h(m_t^2
/M_W^2)}{m_b^2} \left( 1 - \frac{8}{3} \frac{m_c^2}{m_b^2} \right)^{-1}
\simeq 200!
\end{equation}
Here $h(x)$ is a function which
decreases monotonically from 1 at $x=0$ to $1/4$ as $x \to \infty$;
it is about 0.53 for the present value of $m_t$. If the mass difference $\Delta
m / \bar \Gamma \approx 20$ turns out to be too large to measure, the width
difference $\Delta \Gamma / \bar \Gamma \approx 1/10$ may be large enough to
detect.

\section{FLAVOR TAGGING AND SPECTROSCOPY}

In the decays of neutral $B$ mesons to CP eigenstates, it is necessary to know
the flavor of the meson at time of production.  A conventional means for
``tagging'' the flavor of a $B$ is to identify the flavor of the meson produced
in association with it.  At a hadron collider or in high energy $e^+ e^-$
collisions as at LEP, this method suffers only from the possible dilution of
the ``tagging'' signal by $B^0 - \bar B^0$ mixing, and from the difficulty of
finding the ``tagging'' hadron.  However, in the reaction $e^+ e^- \to B^0 \bar
B^0$ at threshold, the odd C of the pair implies that any CP-odd asymmetry will
be an odd function of the time difference between decays of the two mesons,
leading to the need for asymmetric collisions [or, possibly, ingenious schemes
\cite{KB} for enhancing vertex resolution in symmetric machines].

In this section I would like to discuss recent progress in tagging neutral
$B$ mesons by means of the hadrons produced nearby in phase space.  This
method, also proposed \cite{AB} for tagging strange $B$'s via
associated kaons, has been the subject of recent papers devoted
to correlations of nonstrange $B$'s with charged pions \cite{Correls}.

\subsection{Fragmentation vs. Resonances}

The existence of correlations between $B$ mesons and pions can be visualized
either in terms of a fragmentation picture or in terms of explicit resonances.

In a fragmentation picture, if a $b$ quark picks up a $\bar d$ quark from the
vacuum to become a $\bar B^0$ meson, and a charged pion containing the
corresponding $d$ quark is generated, that pion will be a $\pi^-$.  It is
likely to lie ``near'' the $\bar B^0$ in phase space, in the sense that its
transverse momentum with respect to the $B$ is low, its rapidity is correlated
with that of the $B$, or the effective mass of the $\pi B$ system is low.
Similarly, if a $\bar b$ quark picks up a $d$ quark to become a $B^0$, the
charged pion containing the corresponding $\bar d$ will be a $\pi^+$.

The signs of the pions in the above correlations are those which would have
resulted from the decays $B^{**+} \to B^{(*)0} \pi^+$ or $B^{**-} \to \bar
B^{(*)0} \pi^-$. We utilize the double-asterisk superscript to distinguish
$B^{**}$'s from the hyperfine partners of the $B$'s, the $B^*$'s, which are
only about 46 MeV heavier than the $B$'s and cannot decay to them via pions.

The importance of explicit {\em narrow} $B^{**}$ resonances is that they
permit reduction of combinatorial backgrounds.  Thus, we turn to what is
expected (and, more recently, observed) about such resonances.

\subsection{Spectroscopic predictions}

We shall briefly recapitulate material which has been presented in more detail
elsewhere \cite{Charm,Fest}. In a hadron containing a single heavy quark, that
quark ($Q = c$ or $b$) plays the role of an atomic nucleus, with the light
degrees of freedom (quarks, antiquarks, gluons) analogous to the electron
cloud.  The properties of hadrons containing $b$ quarks then can calculated
from the corresponding properties of charmed particles by taking account
\cite{HQS} of a few simple ``isotope effects.''  If $q$ denotes a light
antiquark, the mass of a $Q \bar q$ meson can be expressed as
\begin{equation}
M(Q \bar q) = m_Q + {\rm const.}[n,L] + \frac{\langle p^2 \rangle}{2 m_Q} +
a \frac{\langle {\bf \sigma_q \cdot \sigma_Q} \rangle}{m_q m_Q} + {\cal O}
(m_Q^{-2})~~~.
\end{equation}
Here the constant depends only on the radial and orbital quantum numbers $n$
and $L$.  The $\langle p^2 \rangle /2m_Q$ term expresses the dependence of
the heavy quark's kinetic energy on $m_Q$, while the last term is a hyperfine
interaction.  The expectation value of $\langle {\bf \sigma_q \cdot \sigma_Q}
\rangle$ is $(+1,~-3)$ for $J^P = (1^-,~0^-)$ mesons. If we define
$\overline{M} \equiv [3 M(1^-) + M(0^-)]/4$, we find
\begin{equation}
m_b - m_c + \frac{\langle p^2 \rangle}{2 m_b} - \frac{\langle p^2 \rangle}
{2 m_c} = \overline{M}(B \bar q) - \overline{M}(c \bar q) \simeq 3.34~{\rm
GeV}~~.
\end{equation}
so $m_b - m_c > 3.34$ GeV, since $\langle p^2 \rangle > 0$.  Details of
interest include (1) the effects of replacing nonstrange quarks with strange
ones, (2) the energies associated with orbital excitations, (3) the size of the
$\langle p^2 \rangle$ term, and (4) the magnitude of hyperfine effects.  In all
cases there exist ways of using information about charmed hadrons to predict
the properties of the corresponding $B$ hadrons.  A recent comparison of the
charmed and beauty spectra may be found in Ref.~\cite{Fest}.  For S-wave states
the predictions of the heavy-quark symmetry approach work rather well.

The $B^* - B$ hyperfine splitting scales as the inverse of the heavy-quark
mass: $B^* - B = (m_c/m_b)(D^* - D)$.  Consequently, while $D^{*+} \to D^0
\pi^+$ and $D^{*+} \to D^+ \pi^0$ are both allowed, leading to a useful
method \cite{SN} for identifying charmed mesons via the soft pions often
accompanying them, the only allowed decay of a $B^*$ is to $B \gamma$.  No soft
pions are expected to accompany $B$ mesons.  One must look to the next-higher
set of levels, the $B^{**}$ resonances consisting of a $\bar b$ quark and a
light quark in a P-wave, or the fragmentation process mentioned above, for the
source of pions correlated with the flavor of $B$ mesons.

One can use heavy-quark symmetry or explicit quark models to extrapolate from
the properties of known $D^{**}$ resonances to those of $B^{**}$ states. Two
classes of such resonances are expected \cite{DGG}, depending on whether the
total angular momentum $j = s_q + L$ of the light quark system is 1/2 or 3/2.
Here $s_q$ is the light quark's spin and $L=1$ is its orbital angular momentum
with respect to the heavy antiquark. The light quark's $j = 1/2,~3/2$ can
couple with the heavy antiquark's spin $S_{\bar Q} = 1/2$ to form states with
total angular momentum and parity $J^P_j = 0^+_{1/2},~1^+_{1/2},~1^+_{3/2},
{}~2^+_{3/2}$.

The $0^+_{1/2}$ and $1^+_{1/2}$ states are expected to decay to a ground-state
heavy meson with $J^P = 0^-$ or $1^-$ and a pion via an S-wave, and hence to be
quite broad. No evidence for these states exists in the $\bar c q$ or the $\bar
b q$ system.  By contrast, the $1^+_{3/2}$ and $2^+_{3/2}$ states are expected
to decay to a ground-state heavy meson and a pion mainly via a D-wave, and
hence to be narrow.  Candidates for all the nonstrange and strange $D^{**}$
states of this variety have been identified \cite{Dstars}. The known nonstrange
$D^{**}$ resonances (identified in both charged and neutral states) are a $2^+$
state around 2.46 GeV/$c^2$, decaying to $D \pi$ and $D^* \pi$, and a $1^+$
state around 2.42 GeV/$c^2$, decaying to $D^* \pi$. In addition, strange
$D_s^{**+}$ resonances have been seen, at 2.535 GeV/$c^2$ (a candidate for
$1^+_{3/2}$) and 2.573 GeV/$c^2$ (a candidate for $2^+_{3/2}$). Thus, adding a
strange quark adds about 0.1 GeV/$c^2$ to the mass.

Once the masses of $D^{**}$ resonances are known, one can estimate those of the
corresponding $B^{**}$ states by adding about 3.32 GeV (the quark mass
difference minus a small binding correction).  The results of this exercise are
shown in Table 2.  The reader should consult Ref.~\cite{EHQ} for more detailed
predictions based on potential models and for relations between decay widths of
$D^{**}$ states and those of the $B^{**}$'s.  Thus, the study of excited
charmed states can play a crucial role in determining the feasibility of
methods for identifying the flavor of neutral $B$ mesons.

\begin{table}
\begin{center}
\caption{P-wave resonances of a heavy antiquark and a light quark $q=u,~d$.
In final states $P,~V$ denote a heavy $0^-,~1^-$ meson.  For strange states,
add about 0.1 GeV$/c^2$ to the masses.}
\medskip
\begin{tabular}{c c c c} \hline
$J^P_j$     & $ \bar c q$ mass & $\bar b q$ Mass  &  Allowed final \\
            & (GeV/$c^2$)      & (GeV/$c^2$)      &     state(s)   \\ \hline
$2^+_{3/2}$ &    $2.46^{a)}$   & $\sim 5.77^{b)}$ &  $P \pi, V \pi$ \\
$1^+_{3/2}$ &    $2.42^{a)}$   & $\sim 5.77^{b)}$ &     $V \pi$     \\
$1^+_{1/2}$ &   $<2.42^{c)}$   &  $< 5.77^{b)}$   &     $V \pi$     \\
$0^+_{1/2}$ &   $<2.42^{c)}$   &  $< 5.77^{b)}$   &     $P \pi$     \\ \hline
\end{tabular}
\end{center}
\leftline{$^{a)}$ Observed value.  See Ref.~\protect\cite{Dstars}.}
\leftline{$^{b)}$ Predicted by extrapolation from corresponding $D^{**}$ using
heavy-quark symmetry.}
\leftline{$^{c)}$ Predicted in most quark models.}
\end{table}

\subsection{Spectroscopic observations}

The OPAL \cite{OPB}, DELPHI \cite{DELB}, and ALEPH \cite{ALB} Collaborations
have now observed $B \pi$ correlations which can be interpreted in terms
of the predicted $J^P_j = 1^+_{3/2},~2^+_{3/2}$ states.  The OPAL
data are shown in Fig.~2.  Similar plots have been presented by DELPHI and
ALEPH.  OPAL also sees a $B K$ correlation.

\begin{figure}
\centerline{\psfig{figure=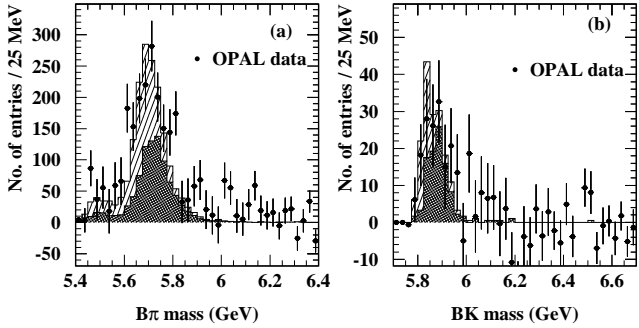,height=2.35in}}
\caption{Evidence for B$^{**}$ resonances obtained by the OPAL Collaboration
\protect\cite{OPB}. (a) $B \pi$ background-subtracted mass distribution; (b)
$B K$ background-subtracted mass distribution.  The solid and cross-hatched
histograms correspond to the contributions of $2^+$ and $1^+$ resonances in
two-resonance fits based on the mass splittings and branching ratios
predicted in Ref.~\protect\cite{EHQ}.}
\end{figure}

In all experiments one is able to measure only the effective mass of a $B \pi$
system.  If a $B^{**}$ decays to $B^* \pi$, the photon in the $B^* \to \gamma
B$ decay is lost, leading to an underestimate of the $B^{**}$ mass by about 46
MeV but negligible energy smearing \cite{Correls}.  Thus, the contributions to
the $B \pi$ mass distribution of $1^+$ and $2^+$ resonances with spacing
$\delta \equiv M(2^+) - M(1^+)$ can appear as three peaks, one at $M(2^+)$ due
to $2^+ \to B \pi$, one at $M(2^+) - 46$ MeV due to $2^+ \to B^* \pi$, and one
at $M(2^+) - \delta - 46$ MeV due to $1^+ \to B^* \pi$.

The OPAL Collaboration fits their $B \pi$ mass distribution either with one
peak with $M = 5681 \pm 11$ MeV/$c^2$ and width $\Gamma = 116 \pm 24$ MeV, or
two resonances, a $1^+$ candidiate at 5725 MeV/$c^2$ with width $\Gamma = 20$
MeV and a $2^+$ candidiate at 5737 MeV/$c^2$ with width $\Gamma = 25$ MeV.  The
widths, mass splittings, and branching ratios to $B \pi$ and $B^* \pi$ in this
last fit are taken from Ref.~\cite{EHQ}, and only the overall mass and
production cross sections are left as free parameters.  The OPAL $B K$ mass
distribution is fit either with a single resonance at $M = 5853 \pm 15$
MeV/$c^2$ with width $\Gamma = 47 \pm 22$ MeV, or two narrow resonances, a
$1^+$ candidate at 5874 MeV/$c^2$ and a $2^+$ candidate at 5886 MeV/$c^2$.

The fitted masses of the nonstrange and strange resonances are respectively
30 MeV lower and 40 MeV higher than the predictions of Ref.~\cite{EHQ}.
The difference could well be due to additional contributions to the
nonstrange channel from the lower-lying $0^+_{1/2}$ and $1^+_{1/2}$ states or
from nonresonant fragmentation. The corresponding strange states might lie
below $BK$ threshold.

The OPAL results imply that $(18 \pm 4)\%$ of the observed $B^+$ mesons are
accompanied by a ``tagging $\pi^-$'' arising from $B^{**0}$ decay.  By isospin
reflection, one should then expect $(18 \pm 4)\%$ of $B^0$ to be accompanied by
a ``tagging $\pi^+$'' arising from $B^{**+}$ decay.  This is good news for the
possibility of ``same-side tagging'' of neutral $B$ mesons.  [Another $(2.6 \pm
0.8)\%$ of the observed $B^+$ mesons are accompanied by a ``tagging $K^-$.''
The isospin-reflected kaon is neutral, and unsuitable for tagging.]

The DELPHI data can be fit with a single peak having a mass of $M = 5732
\pm 5 \pm 20$ MeV and width $\Gamma = 145 \pm 28$ MeV.  The number of
$B^{**}_{u,d}$ per $b$ jet is quoted as $0.27 \pm 0.02 \pm 0.06$. The ALEPH
peak occurs at $M(B \pi) - M(B) = (424 \pm 4 \pm 10)$ MeV/$c^2$, with Gaussian
width $\sigma = (53 \pm 3 \pm 9)$ MeV/$c^2$.  The ALEPH $B^{**}$ signal is
characterized by a production rate
\begin{equation}
\frac{B(Z \to b \to B^{**}_{u,d})}{B(Z \to b \to B_{u,d})} = (27.9 \pm
1.6 \pm 5.9 \pm 3.8)\%~~~,
\end{equation}
where the first error is statistical, the second is systematic, and the
third is associated with uncertainty in ascribing the peak to the contribution
of various resonances.  Multiplying the DELPHI and ALEPH $B^{**}$ rates by the
isospin factor of 2/3 to compare with the OPAL result, we find complete
agreement among the three.

\subsection{Isospin and Correlations}

In principle it should be possible to calibrate the correlations between
charged pions and neutral $B$'s by comparing them with the isospin-reflected
correlations betwen charged pions and {\it charged} $B$'s (whose flavor may be
easier to identify).  Thus, the enhancement of the non-exotic $\pi^+ B^0$
channel with respect to the exotic $\pi^- B^0$ channel should be the same as
that of the non-exotic $\pi^- B^+$ channel with respect to the exotic $\pi^+
B^+$ channel.  What can spoil this relation? I. Dunietz and I \cite{Ispin}
have explored several instances in which care is warranted in making this
comparison.  Some of the differences could be real, but there are many
sources of potential instrumental error against which one has concrete
remedies.

{\it 1. Interaction with the producing system}
can lead to final states which need not be invariant under isospin
reflection.  For example, although a pair of gluons would produce a $b \bar b$
pair with isospin $I = 0$, the fragmentation process could involve picking up
quarks from the producing system (e.g., proton or antiproton fragments) in a
manner not invariant with respect to $u \leftrightarrow d$ substitution.
Similarly, the production of a $B$ meson through diffractive dissociation of
a proton (which has more valence $u$ quarks than $d$ quarks) need not be
invariant under isospin reflection.

{\it 2. Misidentification of associated charged kaons as pions}
can lead one to overestimate the charged $B$ -- charged pion correlations. One
expects $B^+ K^-$ correlations, as seen by OPAL, but not $B^0 K^+$
correlations. As mentioned, the isospin reflection of a charged kaon is
neutral, and would not contribute to a correlation between charged particles
and neutral $B$'s.

{\it 3. Pions in the decay of the associated $B$} will not be produced
in an isospin-reflection-symmetric manner. One must be careful not to
confuse them with primary pions.

{\it 4. Different time-dependent selection criteria} for charged and
neutral $B$'s can lead one to mis-estimate the mixing of neutral $B$'s with
their antiparticles.  (I thank P. Derwent for pointing this out.)  It is
possible to make an unfortunate cut on $B^0$ lifetime which enhances the
mixing considerably with respect to the value obtained by integrating over
all times.

{\it 5. Overestimates of particle identification efficiencies} can lead
to confusion in identification of the flavor of a neutral $B$ through the
decay $B^0 \to J/\psi K^{*0} \to J/\psi K^+ \pi^-$.  It is possible,
especially for $K^+$ and $\pi^-$ with equal laboratory momenta, to confuse
them with $\pi^+$ and $K^-$, while still keeping them in a $K^*$ peak.

The CDF Collaboration at Fermilab has been studying charged pion - $B$
correlations ever since the reports of Ref.~\cite{Correls} appeared, but
no public announcement of these results has yet appeared.  The intent of the
present subsection is {\em not} to provide excuses for this failure to
report results, but rather to provide ideas for ways to view the data in
which such correlations are likely to be robust.  They should certainly
be present in hadron collider data.

\section{DECAYS}

In this section we turn to decays of $B$ mesons to CP non-eigenstates. We have
suggested \cite{BPP} that relations between decays of charged $B$'s
to pairs of light pseudoscalar mesons based on flavor SU(3) \cite{OldSU} could
provide information on weak phases by means of {\em rate measurements alone}.
The latest chapter in this story has been written since the Workshop.

\subsection{$\pi \pi$ and $\pi K$ final states}

Two years ago the CLEO Collaboration \cite{Battle} presented evidence
for a combination of $B^0 \to K^+ \pi^-$ and $\pi^+ \pi^-$ decays, generically
known as $B^0 \to h^+ \pi^-$. On the basis of 2.4 fb$^{-1}$ of data, the most
recent result \cite{Wurt} is $B(B^0 \to h^+ \pi^-) = (1.81^{~+0.6~+0.2}
_{~-0.5~-0.3} \pm 0.2) \times 10^{-5}$.  Although one still cannot conclude
that either decay mode is nonzero at the $3 \sigma$ level, the most likely
solution is roughly equal branching ratios (i.e., about $10^{-5}$) for each
mode.

Other results \cite{CLEOGlas} of the CLEO Collaboration on related modes
include the upper bounds
$B(B^0 \to \pi^0 \pi^0) < 1.0 \times 10^{-5}$,
$B(B^+ \to \pi^+ \pi^0) < 2.3 \times 10^{-5}$, and
$B(B^+ \to K^+ \pi^0) < 3.2 \times 10^{-5}$.
Interesting levels for the last two modes \cite{BPP} are probably around
$(1/2) \times 10^{-5}$, and probably $10^{-6}$ or less for $\pi^0 \pi^0$.
With good particle identification and a factor of several times more data,
it appears that CLEO will be able to make a systematic study of decay modes
with two light pseudoscalars.  What can it teach us?

\subsection{SU(3) relations and the phase $\gamma$}

We mentioned earlier that rate asymmetries in the decays $B^0 \to \pi^+ \pi^-$
and $\bar B^0 \to \pi^+ \pi^-$ could provide information on the weak angle
$\alpha$, as long as a single quark subprocess dominated the decay.  Additional
contributions from penguin diagrams \cite{PP} can be taken into account by
means of an isospin triangle construction \cite{pipi} involving the relation
$A(B^0 \to \pi^+ \pi^-) = \sqrt{2} A(B^0 \to \pi^0 \pi^0) + \sqrt{2} A(B^+
\to \pi^+ \pi^0)$, and the corresponding relation for the charge-conjugate
processes.  Here we define amplitudes such that a partial width is always
proportional to the square of an amplitude with the same coefficient.

A similar amplitude quadrangle applies to the decays $B \to \pi K$ \cite{pik}:
$A(B^+ \to \pi^+ K^0) + \sqrt{2} A(B^+ \to \pi^0 K^+) = A(B^0 \to \pi^- K^+) +
\sqrt{2} A(B^0 \to \pi^0 K^0)$.  When combined with the corresponding relation
for $\bar B^0$ and $B^-$ decays, and used in conjunction with the
time-dependence of the decays $(B^0~{\rm or}~\bar B^0) \to \pi^0 K_S$, these
quadrangles are useful in extracting the weak phase $\alpha$.

In examining SU(3) relations among $B \to PP$ amplitudes, where $P$ is a
pseudoscalar meson, we found that one of the diagonals of the amplitude
quadrangle for $B \to \pi K$ (corresponding to an amplitude with isospin $I =
3/2$) could be related to the purely $I = 2$ amplitude for $B^+ \to \pi^+
\pi^0$.  We obtained the relation
\begin{equation} \label{eqn:tri}
A(B^+ \to \pi^+ K^0) + \sqrt{2} A(B^+ \to \pi^0 K^+) = \tilde r_u \sqrt{2}
A(B^+ \to \pi^+ \pi^0)~~~,
\end{equation}
where $\tilde r_u \equiv (f_K/f_\pi) | V_{us}/V_{ud}|$.  The $B^+ \to \pi^+
K^0$ amplitude is expected to be dominated by a (gluonic) penguin contribution,
involving mainly the combination of CKM elements $V_{tb}^* V_{ts}$, whose
electroweak phase is $\pi$.  The electroweak phase of the $B^+ \to \pi^+ \pi^0$
amplitude is just $\gamma = {\rm Arg}(V_{ub}^*)$.  Thus, in the absence of
strong-interaction phase shift differences, the shape of the amplitude triangle
would give $\gamma$.  One could account for strong-interaction phases by
comparing the amplitude triangle for $B^+$ decays with that for $B^-$ decays
\cite{BPP}. We estimated \cite{JRCP} that in order to measure $\gamma$ to
$10^{\circ}$ one would need a sample including about 100 events in the channels
$\pi^0 K^{\pm}$.

\subsection{Electroweak penguins}

The analyses of Ref.~\cite{BPP} assumed that the only penguin contributions to
$B$ decays were gluonic in nature.  Consequently, one could treat the
flavor-dependence in terms of an effective $\bar b \to \bar d$ or $\bar b \to
\bar s$ transition since the gluon couples to light quarks in a
flavor-symmetric manner.  Thus, the $I = 3/2$ amplitude in $B \to \pi K$
(the diagonal of the $\pi K$ amplitude quadrangle mentioned above) was due
entirely to the Cabibbo-suppressed ``tree-diagram'' process $\bar b \to \bar u
u \bar s$, whose weak phase was well-specified.

It was pointed out \cite{RF,DH} that in certain penguin-dominated $B$ decays
such as $B \to \pi K^*~{\rm and}~\pi K$, electroweak penguin amplitudes were
large enough to compete favorably with the tree amplitude in the $I = 3/2$
channel.  In contrast to gluonic penguins, the virtual photon or $Z$ emitted in
an electroweak penguin diagram does not couple to light quarks in a
flavor-symmetric manner, and possesses an $I = 1$ component. Specifically, if
one decomposes amplitudes into isospin channels,
$$
A(B^+ \to \pi^+ K^0) = (1/3)^{1/2}A_{3/2} - (2/3)^{1/2}A_{1/2}~~,
$$
\begin{equation}
A(B^+ \to \pi^0 K^+) = (2/3)^{1/2}A_{3/2} + (1/3)^{1/2}A_{1/2}~~,
\end{equation}
Deshpande and He \cite{DH} find, in a specific calculation, that
$$
A_{1/2} \sim -0.75 e^{i \gamma} e^{i \delta_{T,1/2}} + 7.3 e^{i \delta_{P,1/2}}
{}~~~,
$$
\begin{equation}
A_{3/2} \sim -1.06 e^{i \gamma} e^{i \delta_{T,3/2}}
+ 0.84 e^{i \delta_{P,3/2}}~~~,
\end{equation}
where the first term in each equation is the ``tree'' contribution (of lowest
order in electroweak interactions), while the second term is the penguin
contribution.  Only the electroweak penguin contributes to the $I = 3/2$
amplitude, but with magnitude comparable to the tree contribution.  The
electroweak penguin spoils the relation of Eq.~(\ref{eqn:tri}).

Recently several of us re-examined the effects of SU(3) breaking \cite{SUbr}
and electroweak penguins \cite{EWP}, to see if one could
extract electroweak penguin effects {\em directly from the data}.
Since our previous SU(3) decomposition gave a complete set of
reduced amplitudes, electroweak penguins only changed the interpretation of
these amplitudes, so that a separation of electroweak penguin effects was
not possible merely on the basis of SU(3).

As pointed out by Deshpande and He \cite{DHP}, certain amplitudes (notably
those for $B_s \to (\pi^0~{\rm or}~\rho^0) + (\eta~{\rm or}~\phi)$ are expected
to be {\rm dominated} by electroweak penguins. We noted that the $\pi K$
amplitude quadrangle could be written in such a manner that one if its
diagonals was equal to $\sqrt{3}A(B_s \to \pi^0 \eta_8)$, where $\eta_8$
denotes an unmixed octet member. The shape of the quadrangle, shown in Fig.~3,
is uniquely determined, up to possible discrete ambiguities. The case of
octet-singlet mixtures in the $\eta$ requires us to replace the $\sqrt{3}$ by
the appropriate coefficient; one can show that the SU(3) singlet contribution
of the $\eta$ is unimportant in this case.

\begin{figure}
\centerline{\psfig{figure=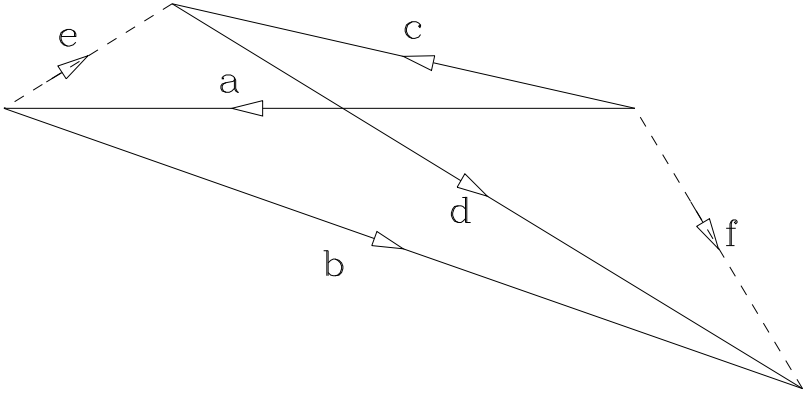,height=2.1in}}
\caption{Amplitude quadrangle for $B \to \pi K$ decays. (a) $A(B^+ \to
\pi^+ K^0)$; (b) $\protect \sqrt 2 A(B^+ \to \pi^0 K^+)$; (c) $\protect
\sqrt 2 A(B^0 \to \pi^0 K^0)$; (d) $A(B^0 \to \pi^- K^+)$; (e) the diagonal
$D_2 = \protect\sqrt 3 A(B_s \to \pi^0 \eta_8)$; (f) the diagonal $D_1 =
A_{3/2}$ corresponding to the $I = 3/2$ amplitude.}
\label{piKquad}
\end{figure}

The quadrangle has been written in such a way as to illustrate the fact
\cite{BPP} that the $B^+ \to \pi^+ K^0$ amplitude receives only penguin
contributions in the absence of ${\cal O}(f_B/m_B)$ corrections. The weak
phases of $\bar b \to \bar s$ penguins, which are dominated by a top quark in
the loop, are expected to be $\pi$. We have oriented the quadrangle to subtract
out the corresponding strong phase, and define corresponding strong phase shift
differences $\tilde \delta$ with respect to the strong phase of the $B^+ \to
\pi^+ K^0$ amplitude.

The $I = 3/2~\pi K$ amplitude is composed of two parts, as noted above.  We
can rewrite it slightly as
\begin{equation}
A_{3/2} = |A_T| e^{i \gamma} e^{i \tilde \delta_{T,3/2}} - |A_P|~~~.
\end{equation}
The corresponding charge-conjugate quadrangle has one diagonal equal to
\begin{equation}
\bar A_{3/2} = |A_T| e^{- i \gamma} e^{i \tilde \delta_{T,3/2}} - |A_P|~~~,
\end{equation}
so that one can take the difference to eliminate the electroweak penguin
contribution:
\begin{equation} \label{eqn:diff}
A_{3/2} - \bar A_{3/2} = 2 i |A_T| \sin \gamma~e^{i \tilde \delta_{T,3/2}}~~~.
\end{equation}
The quantity $|A_T|$ can be related to the $I = 2$ $\pi \pi$ amplitude
in order to obtain $\sin \gamma$. Specifically, if we neglect electroweak
penguin effects in $B^+ \to \pi^+ \pi^0$ (a good approximation), we find that
\begin{equation} \label{eqn:pipi}
|A_T| = \lambda (f_K/f_\pi) \sqrt{2} |A(B^+ \to \pi^+ \pi^0)|~~~.
\end{equation}
Thus, we can extract not only $\sin \gamma$, but also a strong phase shift
difference $\tilde \delta_{T,3/2}$, by comparing Eqs.~(\ref{eqn:diff}) and
(\ref{eqn:pipi}). If such a strong phase shift difference exists, the $B$ and
$\bar B$ quadrangles will have different shapes, and CP violation in the $B$
system will already have been demonstrated.

The challenge in utilizing the amplitude quadrangle in Fig.~3 is to measure
$B(B_s \to \pi^0 \eta)$, which has been estimated \cite{DHP} to be only $2
\times 10^{-7}$!  Very recently (since the Workshop) Deshpande and He
\cite{DHeta} have pointed out that the amplitude triangle
\begin{equation}
2 A(B^+ \to \pi^+ K^0) + \sqrt{2} A(B^+ \to \pi^0 K^+) = \sqrt{6} A(B^+ \to
\eta_8 K^+)~~~,
\end{equation}
implied by the SU(3) relations of Refs.~\cite{BPP,EWP}, where $\eta_8$ is
the octet component of the $\eta$, permits one to specify the
quantity $A_{3/2} - \bar A_{3/2}$ in Eq.~(\ref{eqn:diff}) and extract $\gamma$
as above.  Here there is some delicacy associated with the SU(3) singlet
component of the physical $\eta$.

\subsection{Other final states}

{\it 1. $PV$ final states} ($V = \rho,~\omega,~K^*,~\phi$) are characterized
by more graphs (and hence more reduced SU(3) amplitudes), since one no longer
has the benefit of Bose statistics as in $B \to P P$ decays.  There still
exist quadrangle relations in $\rho K$ and $K^* \pi$ decays, however.
Remarkably, if $\Delta S = 0$ gluonic penguin diagrams (small in any case) are
approximately equal for the cases in which the spectator quark ends up in a
vector meson and in a pseudoscalar [see Ref.~\cite{EWP} for details], the
previous quadrangle construction still holds if we replace $\sqrt{3} A(B_s \to
\pi^0 \eta_8$ with $\sqrt{2} A(B_s \to \pi^0 \phi)$, $A(B \to \pi K)$ with $A(B
\to \pi K^*)$, and $A(B^+ \to \pi^+ \pi^0)$ with $A(B^+ \to \pi^0 \rho^+)$.

Deshpande and He \cite{DHP} predict $B(B_s \to \pi^0 \phi) \approx 2 \times
10^{-8}$, which probably means that the quadrangle reduces to two nearly
overlapping triangles (whose shapes will consequently be difficult to specify).
On the other hand, in $PV$ decays, the effects of electroweak penguins then
may not be so important if the dominant processes are characterized by
branching ratios of order $10^{-5}$ as in $B \to PP$ decays.

Hints of some signals have been seen in some $PV$ channels in the latest
CLEO data \cite{Wurt}, but only upper limits are being quoted.  These are
fairly close to theoretical expectations in the case of some $\pi \rho$
channels.

{\it 2. $VV$ final states} satisfy Bose statistics. Since the total angular
momentum of the decaying particle is zero, the (space) $\times$ (spin) part of
the $VV$ wave function will be symmetric, as in $PP$ final states \cite{HJLpc}.
Thus, there should exist amplitude relations for each relative orbital angular
angular momentum $\ell$.  If one $\ell$ value dominates the decays, such
relations might be tested using triangles constructed of square roots of decay
rates, as in the $PP$ case.

\section{SPECIFICS FOR HADRON COLLIDERS}

In this section we comment on miscellaneous issues which deserve more attention
if hadron collider experiments are to make a dent in the problem of CP
violation in the $B$ meson system.

\subsection{Vertex resolution}

As we mentioned earlier, if different minimum decay lengths are adopted for
selection of different $B^0$ decay modes, the data sample will contain
different admixtures of $B^0$ and $\bar B^0$ in different decay modes. Apparent
inconsistencies can show up in charged pion -- neutral $B$ correlations.

A trigger based on the presence of secondary vertices may
be the only way to pick out elusive decays such as $B \to \pi \pi$ in sufficent
numbers for CP studies.

\subsection{Particle Identification}

In the comparison of $B \to \pi \pi$ and $B \to \pi K$ decay modes, we have
seen that good discrimination between pions and kaons is essential.  Some
distinctive processes of $B$ production and decay involve large number of
kaons.  Consider, for example the production of a $B_s = \bar b s$ with a kaon
``tag.''  If the subsequent decay is $B_s \to D_s^- K^+ \to \phi \pi^- K^+ \to
K^+ K^- \pi^- K^+$, a total of four kaons in the final state (counting the
``tagging'' kaon) must be identified.

\subsection{Particle-Antiparticle Correlations}

In the production of a $b \bar b$ pair in high-energy $e^+ e^-$ collisions, as
at LEP, the $b$ and $\bar b$ are separated from one another by many units of
rapidity, with many hadrons filling the gap in between.  In hadronic $b \bar
b$ production, there may be a continuum of rapidity
separations between $b$ and $\bar b$, ranging from the very small (as in
fragmentation of a hard gluon to $b \bar b$) to the very large (as in
back-to-back production of $b$ and $\bar b$ jets). The ``same-side'' tagging
method is likely to be more relevant for widely separated $b$ and $\bar b$.
One must learn more about the
hadrons which fill the gap between a $b$ and a $\bar b$ in hadronic collisions,
and tailor one's tagging methods accordingly.

\section{CONCLUSIONS}

The most promising advantage for the study of $B$ physics at the Tevatron
(and ultimately, at the LHC) consists of the high rate of $b$ production.
The relatively high
$b \bar b$ effective mass expected in such processes implies that ``same-side''
tagging of neutral $B$ mesons in order to identify CP-violating rate
asymmetries
deserves serious study. A study of lifetime differences of hadrons containing
$b$ quarks at an accuracy of several percent will permit the checking of
detailed predictions of these differences.

The innate advantage (based on rate) of the Fermilab detectors for
identification of tagged neutral $B$ mesons became clear
several years ago when the first reconstructed $B$ mesons were presented by the
CDF Collaboration.  However, ``same-side'' tagging
results from CDF have been nearly two years in gestation.
Meanwhile, the recent identification of ``$B^{**}$'' mesons at LEP has brought
to mind the last of Satchel Paige's six rules for staying young \cite{Paige}:
\begin{quote}
Don't look back. Something may be gaining on you.
\end{quote}

\section*{ACKNOWLEDGMENTS}

I wish to thank M. Gronau, O. F. Hern\'andez, and D. London for enjoyable
collaborations on some of the topics menioned here; C. Quigg and M. Shochet for
constructive comments; F. Bedeschi, D. Cassel, M. Feindt, R. Fleischer, R.
Forty, L. K. Gibbons, R. Kowalewski, H. Lipkin, S. Schael, P. Sphicas, and F.
W\"urthwein for helpful correspondence; P. Derwent and D. Gerdes for help with
the last reference; and M. Worah for comments on the manuscript. This work was
supported by the United States Department of Energy under Grant No.~DE FG02
90ER40560.

\def \ajp#1#2#3{Am.~J.~Phys.~{\bf#1}, #2 (#3)}
\def \apny#1#2#3{Ann.~Phys.~(N.Y.) {\bf#1}, #2 (#3)}
\def \app#1#2#3{Acta Phys.~Polonica {\bf#1}, #2 (#3)}
\def \arnps#1#2#3{Ann.~Rev.~Nucl.~Part.~Sci.~{\bf#1}, #2 (#3)}
\def \baps#1#2#3{Bull.~Am.~Phys.~Soc.~{\bf#1}, #2 (#3)}
\def \cmts#1#2#3{Comments on Nucl.~Part.~Phys.~{\bf#1}, #2 (#3)}
\def \cn{Collaboration}
\def \cp89{{\it CP Violation,} edited by C. Jarlskog (World Scientific,
Singapore, 1989)}
\def \dpff{{\it The Fermilab Meeting DPF 92} (7th Meeting of the American
Physical Society Division of Particles and Fields, 10 -- 14 November 1992,
Fermi National Accelerator Laboratory), edited by C. H. Albright \ite~(World
Scientific, Singapore, 1993)}
\def \efi{Enrico Fermi Institute Report No.~EFI~}
\def \f79{{\it Proceedings of the 1979 International Symposium on Lepton and
Photon Interactions at High Energies,} Fermilab, August 23-29, 1979, ed. by
T. B. W. Kirk and H. D. I. Abarbanel (Fermi National Accelerator Laboratory,
Batavia, IL, 1979}
\def \hb87{{\it Proceeding of the 1987 International Symposium on Lepton and
Photon Interactions at High Energies,} Hamburg, 1987, ed. by W. Bartel
and R. R\"uckl (Nucl. Phys. B, Proc. Suppl., vol. 3) (North-Holland,
Amsterdam, 1988)}
\def \ib{{\it ibid.}~}
\def \ibj#1#2#3{~{\bf#1}, #2 (#3)}
\def \ichep72{{\it Proceedings of the XVI International Conference on High
Energy Physics}, Chicago and Batavia, Illinois, Sept. 6 -- 13, 1972,
edited by J. D. Jackson, A. Roberts, and R. Donaldson (Fermilab, Batavia,
IL, 1972)}
\def \ijmpa#1#2#3{Int. J. Mod. Phys. A {\bf#1}, #2 (#3)}
\def \lkl87{{\it Selected Topics in Electroweak Interactions} (Proceedings of
the Second Lake Louise Institute on New Frontiers in Particle Physics, 15 --
21 February, 1987), edited by J. M. Cameron \ite~(World Scientific, Singapore,
1987)}
\def \ky85{{\it Proceedings of the International Symposium on Lepton and
Photon Interactions at High Energy,} Kyoto, Aug.~19-24, 1985, edited by M.
Konuma and K. Takahashi (Kyoto Univ., Kyoto, 1985)}
\def \mpla#1#2#3{Mod. Phys. Lett. A {\bf#1}, #2 (#3)}
\def \nc#1#2#3{Nuovo Cim. {\bf#1}, #2 (#3)}
\def \np#1#2#3{Nucl. Phys. {\bf#1}, #2 (#3)}
\def \pisma#1#2#3#4{Pis'ma Zh. Eksp. Teor. Fiz. {\bf#1}, #2 (#3) [JETP Lett.
{\bf#1}, #4 (#3)]}
\def \pl#1#2#3{Phys. Lett. {\bf#1}, #2 (#3)}
\def \plb#1#2#3{Phys. Lett. B {\bf#1}, #2 (#3)}
\def \pr#1#2#3{Phys. Rev. {\bf#1}, #2 (#3)}
\def \prd#1#2#3{Phys. Rev. D {\bf#1}, #2 (#3)}
\def \prl#1#2#3{Phys. Rev. Lett. {\bf#1}, #2 (#3)}
\def \prp#1#2#3{Phys. Rep. {\bf#1}, #2 (#3)}
\def \ptp#1#2#3{Prog. Theor. Phys. {\bf#1}, #2 (#3)}
\def \rmp#1#2#3{Rev. Mod. Phys. {\bf#1}, #2 (#3)}
\def \rp#1{~~~~~\ldots\ldots{\rm rp~}{#1}~~~~~}
\def \si90{25th International Conference on High Energy Physics, Singapore,
Aug. 2-8, 1990}
\def \slc87{{\it Proceedings of the Salt Lake City Meeting} (Division of
Particles and Fields, American Physical Society, Salt Lake City, Utah, 1987),
ed. by C. DeTar and J. S. Ball (World Scientific, Singapore, 1987)}
\def \slac89{{\it Proceedings of the XIVth International Symposium on
Lepton and Photon Interactions,} Stanford, California, 1989, edited by M.
Riordan (World Scientific, Singapore, 1990)}
\def \smass82{{\it Proceedings of the 1982 DPF Summer Study on Elementary
Particle Physics and Future Facilities}, Snowmass, Colorado, edited by R.
Donaldson, R. Gustafson, and F. Paige (World Scientific, Singapore, 1982)}
\def \smass90{{\it Research Directions for the Decade} (Proceedings of the
1990 Summer Study on High Energy Physics, June 25--July 13, Snowmass,
Colorado),
edited by E. L. Berger (World Scientific, Singapore, 1992)}
\def \stone{{\it $B$ Decays}, Revised 2nd Edition, edited by S. Stone (World
Scientific, Singapore, 1994)}
\def \tasi90{{\it Testing the Standard Model} (Proceedings of the 1990
Theoretical Advanced Study Institute in Elementary Particle Physics, Boulder,
Colorado, 3--27 June, 1990), edited by M. Cveti\v{c} and P. Langacker
(World Scientific, Singapore, 1991)}
\def \yaf#1#2#3#4{Yad. Fiz. {\bf#1}, #2 (#3) [Sov. J. Nucl. Phys. {\bf #1},
#4 (#3)]}
\def \zhetf#1#2#3#4#5#6{Zh. Eksp. Teor. Fiz. {\bf #1}, #2 (#3) [Sov. Phys. -
JETP {\bf #4}, #5 (#6)]}
\def \zpc#1#2#3{Zeit. Phys. C {\bf#1}, #2 (#3)}

\end{document}